\documentclass[a4paper,11pt, draftcls, onecolumn]{IEEEtran}
\usepackage[latin1]{inputenc}
\usepackage[english]{babel}
\usepackage[T1]{fontenc}
\usepackage{array}
\usepackage{graphicx}
\usepackage[caption=false,font=footnotesize]{subfig}
\begingroup\expandafter\expandafter\expandafter\endgroup
\expandafter\ifx\csname IncludeInRelease\endcsname\relax
  \usepackage{fixltx2e}
\fi
\usepackage{cite}
\usepackage{algorithm,algorithmicx,algpseudocode}
\usepackage{xfrac}
\usepackage{wrapfig}    
\usepackage[cmex10]{amsmath}
\usepackage{amsfonts,amssymb}
\usepackage{booktabs}
\usepackage{color}
\usepackage{epstopdf}
\usepackage{float}
\usepackage[T1]{fontenc}
\usepackage[cmex10]{amsmath}
\usepackage{amsfonts,amssymb}
\usepackage{multicol}
\usepackage{url}
\usepackage{yfonts}
\usepackage{tikz}
\definecolor{gold}{rgb}{0.85,.66,0}

\begin{document}
% ============================================
\title{DE/PSO-aided Hybrid Linear Detectors for MIMO-OFDM Systems under Correlated Arrays}
% ============================================
\author{{Rafael Masashi Fukuda}, David William Marques Guerra, Ricardo Tadashi Kobayashi, Taufik Abrao \vspace{-5mm}}

\maketitle

\begin{abstract}
In this paper, we analyze the performance of evolutionary heuristic-aided linear detectors deployed in Multiple-Input Multiple-Output (MIMO) Orthogonal Frequency-Division Multiplexing (OFDM) systems, considering realistic operating scenarios. Hybrid linear-heuristic detectors under different initial solutions provided by linear detectors are considered, namely differential evolution (DE) and particle swarm optimization (PSO). Numerical results demonstrated the applicability of hybrid detection approach, which can improve considerably the performance of minimum mean-square error (MMSE) and matched filter (MF) detectors. Furthermore, we discuss how the complexity of the presented algorithms scales with the number of antennas, besides of verifying the spatial correlation effects on MIMO-OFDM performance assisted by linear, heuristic and hybrid detection schemes. The influence of the initial point in the performance improvement and complexity reduction is evaluated numerically.
\end{abstract}


\IEEEpeerreviewmaketitle

%-=================
\section{Introduction}
%-=================
Contemporary wireless communications systems, such as IEEE 802.11 and 4G LTE, deploy multicarrier modulation with the aim of transmitting data over frequency-selective channels. In this sense, OFDM is the most popular choice and a suitable number of subcarriers is used to make subchannels frequency flat. Moreover, dispersion and other phenomena introduce undesirable effects that may limit the overall performance of a wireless system. From this perspective, authors in \cite{Saeed2003} discuss how the number of subcarriers affects the transmission of an OFDM signal with equipped with a single antenna at both transmission sides transmitter-receiver (SISO). 

In the search of more efficient systems, Multiple-input multiple-output (MIMO) systems were proposed and able to improve the spectral efficiency \cite{Hampton:2014}. However, such benefits also require more sophisticated electrical circuitry and signal processing, which are needed to decouple signals from the different antennas \cite{Goldsmith2005}. The system may increase the throughput using multiplexing mode, where each antenna transmit different signals. Conversely, increasing the performance/reliability requires the transmission of the same information and exploiting diversity. Those characteristics are limited to the Diversity-Multiplexing Tradeoff \cite{Tse2004}. Herein, the multiplexing mode is considered, where the signal of the other $N_t - 1$ transmit antennas interfere each other. Thus, detection algorithms are required to reduce the effects of such interference \cite{Choi:2012},\cite{Kobayashi2015} and are studied throughout this work. 

In order to attain high levels of efficiency, the MIMO system considers the assumption of rich scattering (isotropic) scenario modeled as independent Rayleigh \cite{marzetta2016}, 
which is not always entirely valid in real applications. A rule of thumb is the approximation of half wavelength of separation between antennas \cite{Goldsmith2005} to achieve independent fading channels, but this distance may not be always respected, for example, due to space limitation of the receiver hardware, resulting in spatial correlation of the channel coefficients. In realistic scenarios, correlated models are good representations of field measurements \cite{Chizhik2003}, and thus considered in our numerical simulations. 

Authors in \cite{Guerra2016} discuss how the performance of SISO-OFDM systems scale with the number of subcarriers. In the MIMO-OFDM context, the performance of ZF and MMSE linear detectors are analyzed under spatial correlation scenarios. This work extends the results reported in \cite{Guerra2016}. 
In particular and differently of \cite{Guerra2016}, herein, we propose a hybrid detection approach, where particle swarm optimization (PSO) and  differential evolution (DE) evolutionary heuristics are combined with linear detectors (two detection steps), aiming to improve performance with reduced  increment in complexity.

In detection problem, the maximum likelihood (ML) is known to provide optimal performance, however its high computational complexity is prohibitive in real applications, specially when the problem dimension increases, e.g., number of antennas, constellation size and number of subcarriers. Heuristic algorithms provide alternative good solutions with relatively low computational complexity. In \cite{Khan2006}, PSO-aided detection is considered in MIMO and in \cite{trimeche2013} to MIMO-OFDM systems, providing lower computational complexity compared to ML detector. In \cite{Seyman_2014}, heuristic approaches differential evolution (DE), genetic algorithm (GA) and PSO are applied to detection in MIMO-OFDM and performance in terms of bit error rate (BER) is evaluated. In \cite{Khan2007}, binary PSO (BPSO) is applied to MIMO-OFDM and an algorithm considering the output of ZF-VBLAST is proposed and performance evaluated numerically. 

The contributions of this paper are as follows. We analyse the influence on BER performance and computational complexity in terms of \textit{floating points operations} (\textsc{flop}s) of different initial solution as input to the heuristic algorithms, {\it i.e.},we have analyzed distinct initialization, including random guess, linear detector outputs, such as MF and MMSE solutions as input, while perform a comparison between those heuristic detectors in realistic scenario, {\it i.e.}, under spatial correlation between antennas. Moreover, aiming to attain a fair performance-complexity comparison, the input parameters of both heuristic strategies have been systematically chosen, since they directly impact on the algorithm performance and complexity, as studied in \cite{Marinello_2012}.

The remainder of this work is organized as follows. Section II revisits briefly the OFDM scheme. Descriptions for the MIMO-OFDM system with spatial channel correlation are offered in section \ref{sec:mimo}. Moreover, section \ref{sec:detectors} also describes the classical MIMO detectors and formulates heuristic aided detectors based on PSO and DE, including the hybrid linear-heuristic approaches. Extensive numerical results are discussed in section \ref{sec:simulation_results}, where BER performance comparison considering spatial correlation was systematically carried out. Besides, subsection \ref{ref:complexity} carefully analyzes the resulting complexity of the MIMO-OFDM detectors. Final remarks and conclusions are offered in section \ref{sec:conclusions}.

\noindent{\it Notation}: Throughout the paper, lowercase and uppercase bold-faced letters represent vectors and matrices, respectively. $\mathbb{C}$ and $\mathbb{R}$ the set of complex and real numbers; $\mathfrak{Re\{.\}}$ and $\mathfrak{Im\{.\} }$ represent the real and imaginary parts of a complex number. Operators $[.]^H$, $\|.\|$, $\circ$ and $\otimes$ represent Hermitian, Frobenius norm, Hadamard product and Kronecker product, respectively. $\mathbb{E}\{.\}$ denotes expectation operator and and $\sim \mathcal{U} \in [a, b]$  that a random variable follows an uniform distribution inside a specified interval.

%===========================================================================================
\section{OFDM Transmission and Channel}\label{sec:ofdm}
\begin{figure}[htbp!]
\centering
\includegraphics[width=.5\textwidth]{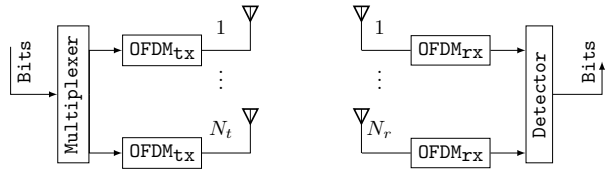}
\vspace{-2mm} 
\caption{MIMO-OFDM block diagram.}
\label{fig:mimo_ofdm_block_diagram}
\end{figure}

%===========================================================================================
A block diagram representing the MIMO-OFDM communication in multiplexing operation mode is exposed in Fig. \ref{fig:mimo_ofdm_block_diagram}. At the transmitter side, the stream of bits are distributed throughout $N_t$ transmitting substreams. Here, classical OFDM modulation is considered and described as follows. The signal passes through the $OFDM_{\rm tx}$ block that represents the OFDM modulator, which includes the serial-to-parallel conversion, digital $M$-ary modulation, inverse discrete Fourier transform (IDFT), cyclic prefix (CP) addition, parallel-to-serial conversion and the transmission of the signal through the wireless channel. At the receiver, the signals of the $N_r$ receive antennas are shifted to baseband, passed by the OFDM demodulator ($OFDM_{\rm rx}$), which includes a serial-to-parallel followed by a discrete Fourier transform (DFT). Thus, CP is discarded, the signal is serialized, demodulated and it finally feeds the detection block, which is the focus of this work. Note that linear, heuristic and hybrid detectors are discussed in more details in section \ref{sec:detectors}.

Among the different channel effects, the coherence time $(\Delta t)_c$ and the coherence band $(\Delta B)_\textsc{c}$ may influence parameters of an OFDM system. The coherence time scales directly with the maximum Doppler frequency while the mobility of a wireless terminal may cause problems such as the {\it carrier frequency offset} \cite{Cho2010}, which is important for the performance of the system but not the focus of this paper. 
The coherence bandwidth is dictated by power delay profile (PDP) of the channel, which is measured empirically \cite{Goldsmith2005}. More specifically, the the coherence bandwidth is evaluated based on the estimation of the delay spread of the PDP of a channel.
This parameter influences directly on the number of subcarriers of the system, because, to achieve the flat-fading on every subchannel, the condition $B_{\rm sc} \ll (\Delta B)_\textsc{c}$ requires $N$ to be sufficiently large \cite{Goldsmith2005}. In special, this work deploys the IEEE 802.11b PDP model, which follows an exponential profile \cite{Cho2010}.

%======================================================================================
\section{MIMO-OFDM Multiplexing Mode and Spatial Correlation}\label{sec:mimo}
%======================================================================================
Considering $N_t$ and $N_r$ transmit and receive antennas, respectively, the signal received in a MIMO-OFDM channel on each subcarrier can be expressed as \cite{Paulraj2003}:
\begin{equation}\label{eq:mimo}
{\bf y}[n]={\bf H}[n] {\bf x}[n] + {\bf z}[n],
\end{equation}
where ${\bf y}[n] \in \mathbb{C}^{N_r\times 1} $ is the vector of the received signal, ${\bf H}[n]  \in \mathbb{C}^{N_r \times N_t}$ is the channel matrix, ${\bf x}[n]  \in \mathbb{C}^{N_t\times 1}$ the transmitted information, ${\bf z}[n] \in \mathbb{C}^{N_r\times 1}$ the Gaussian noise with zero mean and variance $\sigma_z^2$ through $n=0, \cdots, N-1$ subcarriers.

%-------------------
In order to describe and evaluate spatial correlation between antennas, the Kronecker product is used as follows:
\begin{eqnarray}
{\bf H}[n] = \sqrt{{\bf R}_r}{\bf G}[n] \sqrt{{\bf R}_t^{H}}, 
\end{eqnarray}
where $\bf G$ is an uncorrelated channel matrix composed by independent and identically distributed (i.i.d.) entries, ${{\bf R}_r}$ and ${{\bf R}_t}$ are the spatial correlation matrices seen by the receiver and transmitter, respectively. 
The coefficients needed to construct the correlation matrix and the arrange of the antennas (linear, rectangular) influence the entries of correlation matrices of the transmitter and receiver. 

In \cite{Zelst2002}, an antenna correlation model is proposed for {\it uniform linear antenna} (ULA) array configurations.  This model considers that the antennas are arranged equidistantly, where $d_t$ and $d_r$ represent the spacing between the transmitting and receiving antennas, linearly arranged, respectively.  To simplify the analysis, we consider $N_t = N_r$, leading to Toeplitz symmetric correlation matriz:
\begin{eqnarray}
{\bf R}_t = {\bf R}_r =
\begin{bmatrix}
1     & \rho  & \rho^4 & \dots & \rho^{(N_t-1)^2} \\
\rho  & 1     &       &       & \vdots \\
\rho^4 & \rho  & 1     &       & \rho^4 \\
\vdots & \vdots & \vdots & \ddots & \rho \\
\rho^{(N_t-1)^2} & \dots & \rho^4 & \rho  & 1 \\
\end{bmatrix}
,
\end{eqnarray}
where $\rho\in [0,\,1]$ denotes the correlation index between element antennas of a ULA array.

%===================================
\section{MIMO-OFDM Detectors}\label{sec:detectors}
%===================================
In this section, linear and heuristic-based detectors are discussed in details. Heuristic procedure involves the definition of a fitness function, deployed to evaluate the quality of the population/swarm and to decide which ones are more suitable to solve a given problem (in this paper, MIMO-OFDM detection). Furthermore, the model is rewritten in an equivalent real-valued representation and the PSO and DE heuristic procedures are detailed, while the utilization of different initial solution (hybrid approach) is briefly described.

%================================
\subsection{Maximum likelihood (ML) Detector}
%================================
Aiming to perform optimal symbol estimation, ML detection requires an exhaustive search over all symbol vector combinations. However, optimal performance comes at high computational complexity, which is not feasible for real world systems. In the search, the vector that offers the minimum Euclidean distance between the actual received signal ${\bf y}[n]$ and the estimated reconstructed received signal ${\bf H}[n]{\bf x}[n]$, assuming the transmission of a given candidate-signal vector ${\bf x}[n]$. Hence, ML symbols estimation for MIMO-OFDM systems can be formulated as the following problem:
\begin{equation}
\tilde{\bf x}[n] = \min_{\bf x} \| {\bf y}[n] - {\bf H}[n] {\bf x}[n]\|^2.
\end{equation}

%===========================================
\subsection{Linear Detectors}
%===========================================
Since MIMO channels introduce linear superposition between the transmitted signals, detection algorithms must be deployed at the receiver side to mitigate inter-antenna interference while allow the symbol reconstruction \cite{Cho2010}. In this sense, the ZF is one of the simplest MIMO-OFDM equalizers which uses the Moore-Penrose pseudo-inverse matrix to decouple the transmitted symbol vector, i.e.:
\begin{equation}\label{eq: ZFD}
{\bf H}^\dagger_{\rm zf}[n] = ({\bf H}[n]^H {\bf H}[n])^{-1} {\bf H}[n]^H.
\end{equation}

Alternatively, the MMSE linear detector considers the statistical distribution of the noise. Therefore, this detector aims to minimize the distance between the the actual transmitted signal and the estimated signal obtained through a linear equalization matrix \cite{Hampton:2014}. Such optimization procedure can be defined by
\begin{equation}
{\bf H}^\dagger_{\rm mmse}[n]  =  \min_{\bf W}\,\,  \mathbb{E}\left\{\|\boldsymbol{\mathrm{x}}[n] - {\bf Wy}[n]\|^2\right\}.\label{eq:mmsea}
\end{equation}
Thus, solving eq. \eqref{eq:mmsea} leads to the MMSE closed form solution 
\begin{equation}\label{eq:MMSED}
{\bf H}^\dagger_{\rm mmse}[n] = \left({\bf H}^H[n] {\bf H}[n] + \dfrac{N_0}{E_S}{\bf I} \right)^{-1}{\bf H}^H[n].
\end{equation}
where $\frac{N_0}{E_S}$ is the inverse of the signal-to-noise ratio (SNR).

As another option, the matched filter (MF) is a classical method that provides optimum performance in the AWGN scenario, and consists of the multiplication of the received signal by the transpose conjugate of the channel.

Finally, linear estimation can be generically described by
\begin{equation}\label{eq:linearDetectors}
\tilde{\bf x}[n] = {\bf W}_{\rm lin}[n] \,{\bf y}[n],
\end{equation}
where ${\bf W}_{\rm lin}[n] = {\bf H}^\dagger_{\rm zf}[n]$ for the ZF detection, ${\bf W}_{\rm lin}[n] = {\bf H}^\dagger_{\rm mmse}[n]$ for the MMSE detection and ${\bf W}_{\rm lin} = {\bf H}^{H}[n]$ for the matched filter. 

%======================================
\subsection{Fitness Function} 
%======================================
To facilitate the application of the heuristic methods, eq.\eqref{eq:mimo} can be denoted as an equivalent real-valued representation as follows:
\begin{equation}
{\underline{\bf y}[n]} = 
\begin{bmatrix}
\mathfrak{Re}\{ {\bf y}[n] \}     \\  \mathfrak{Im}\{ {\bf y}[n] \}
\end{bmatrix}
,\quad 
\underline{\bf H}[n] = 
\begin{bmatrix}
\mathfrak{Re}\{ {\bf H}[n] \}     &  -\mathfrak{Im}\{ {\bf H}[n] \}     \\
\mathfrak{Im}\{ {\bf H}[n] \}     &  \mathfrak{Re}\{ {\bf H}[n] \}
\end{bmatrix} 
, 
\end{equation}
\begin{equation}
\qquad {\underline{\bf x}[n]} = 
\begin{bmatrix}
\mathfrak{Re}\{ {\bf x}[n] \}     \\  \mathfrak{Im}\{ {\bf x}[n] \}
\end{bmatrix}
, \qquad {\underline{\bf z}[n]} = 
\begin{bmatrix}
\mathfrak{Re}\{ {\bf z}[n] \}     \\  \mathfrak{Im}\{ {\bf z}[n] \}
\end{bmatrix}
,
\end{equation}
where matrix ${\underline{\bf H}} \in \mathbb{R}^{2N_r \times 2N_t}$ and vectors $\underline{\bf y}[n] \in \mathbb{R}^{2N_r \times 1}, \underline{\bf x}[n]$ and $\underline{\bf z}[n] \in \mathbb{R}^{2N_t \times 1} $ are the real-valued representation of the channel, received signal, sent information and thermal noise, respectively.

For the detection problem, generally, the fitness function is defined based on the Euclidean distance between the received signal and the estimated-reconstructed (candidate) symbol, and formulated as \cite{trimeche2013,Seyman_2014,Khan2007}:
\begin{equation}\label{eq:fitness}
f( \boldsymbol{\zeta} ) = \|\underline{\bf y}[n]-\underline{\bf H}[n]\boldsymbol{\zeta} \|^2.
\end{equation}
where $\zeta$ denotes the entity that we want to evaluate, an specific position of particle in PSO and an individual in DE.

%======================================
\subsection{Heuristic PSO-based Detector}
%======================================
PSO is an evolutionary heuristic algorithm with adjustable parameters, such as cognitive and social factors ($c_1$ and $c_2$ respectively), related to the behavior of bird flocking and fish schooling. Associated to each particle there is a velocity ${\bf v}\in \mathbb{R}^{N_{\rm dim}\times 1}$, actual position ${\bf p}\in \mathbb{R}^{N_{\rm dim}\times 1}$ and personal best position ${\bf p}_{\textsc{pb}}\in \mathbb{R}^{N_{\rm dim}\times 1}$ associated, that are updated at each iteration of the algorithm as follows in matrix representation \cite{Cheng2011}: 
\begin{equation}\label{eq:psoVelocity}
{\bf V} = w{\bf V} + c_1 {\bf U}_1 \circ ({\bf M}_{\textsc{pb}}{\bf - P}) + c_2 {\bf U}_2 \circ ( {\bf M}_{\textsc{gb}}{\bf - P} ),
\end{equation}
\begin{equation}\label{eq:psoPostion}
\bf P = P + V,
\end{equation}
where $N_{\rm dim}$ denotes the dimensionality of the problem, $w$ the inertia factor; ${\bf U}_1$ and ${\bf U}_2$ are matrices compounded of elements $\sim \mathcal{U}[0,1]$, ${\bf P}\in \mathbb{R}^{N_{\rm dim}\times N_{\rm pop}}$ and ${\bf V} \in \mathbb{R}^{N_{\rm dim}\times N_{\rm pop}}$ matrices store the position and velocity of $N_{\rm pop}$ particles of the swarm in each column, i.e., ${\bf P} = [{\bf p}_{1} \dots {\bf p}_{N_{\rm pop}}]$ and ${\bf V} = [{\bf v}_{1} \dots {\bf v}_{N_{\rm pop}}]$. ${\bf M}_{\textsc{pb}}$ is a matrix constructed with the personal best position of each particle and the best position matrix is given by ${\bf M}_{\textsc{gb}} = [{\bf p}_{\textsc{gb}} \dots {\bf p}_{\textsc{gb}}] \in \mathbb{R}^{N_{\mathrm{dim}} \times N_{\mathrm{pop}} }$, where vector ${\bf p}_{\textsc{gb}}\in \mathbb{R}^{N_{\rm dim}\times 1}$ denotes the best position in the swarm, the global best (in a minimization problem, the position that provides the lowest value of the fitness function).

The $w$ coefficient introduced in \cite{Shi1998} can be a constant, linear or nonlinear function and it balances the global and local exploitation depending on its value \cite{Shi1998b}. Here, a nonlinear decreasing strategy of $0.99w$ is considered. Regarding the velocity, to avoid a possible increase to infinity, it was limited to the interval $[-V_{\textsc{max}}, V_{\textsc{max}}]$ \cite{Shi1998b}, with $V_{\textsc{max}}$ representing the maximum possible velocity value.

After the execution of $N_{\rm iter}$ times of the PSO algorithm, the output vector ${\bf p}_{\textsc{gb}}$ corresponds to the detected symbol using the PSO-aided detector $\tilde{\bf x}_{\textsc{pso}}[n]$ in the MIMO-OFDM problem.

\begin{algorithm}[h]
\caption{ PSO -- Particle Swarm Optimization.}\label{algo:PSO}
\begin{algorithmic}[1]
\small
\State{ Input parameters: \, $c_1, c_2, w, N_{\mathrm{pop}}, N_{\mathrm{iter}}, \bf P$}
\State{Initialization of ${\bf M}_{\textsc{pb}}$ and ${\bf M}_{\textsc{gb}}$}
\For {$1 \to  N_{\mathrm{iter}}$}
\State{ Calculate velocity, eq. \eqref{eq:psoVelocity}} 
\State{ Calculate position, eq. \eqref{eq:psoPostion}}
\State{ Evaluate fitness function, eq. \eqref{eq:fitness}, for all particles }
\State{ Update personal best matrix ${\bf M}_{\textsc{pb}}$}
\State{ Update global best matrix ${\bf M}_{\textsc{gb}}$}
\EndFor
\State{Output: \, ${\bf p}_{\textsc{gb}}$}
\end{algorithmic}
\end{algorithm}

%===================================
\subsection{Heuristic DE-based Detector}
%===================================
DE is a population-based heuristic proposed in \cite{Storn1997} that relies on operations mutation, crossover and selection in order to avoid be trapped on local minima across the $N_{\rm gen}$ generations of the algorithm.

Consider $\boldsymbol{\iota}, \boldsymbol{\nu} , \boldsymbol{\psi}$  vectors with dimensions $N_{\rm dim}\times 1$ that represent the individuals, mutation and crossover vectors, while $N_{\rm ind}$ is the number of individuals. The operations of the DE algorithm operating with the strategy \texttt{rand/1/bin} presented in \cite{Storn1997} are synthesized in the following.

%-------------------------------------------------
\subsubsection{Mutation}
%-------------------------------------------------
At each iteration, the $k-$th mutation vector is constructed as:
\begin{equation}\label{eq:mutation}
\boldsymbol{\nu}_{k} = \boldsymbol{\iota}_{r_1} + F_{\rm mut}(\boldsymbol{\iota}_{r_2} - \boldsymbol{\iota}_{r_3}),
\end{equation}
where variables $r_1 \neq r_2 \neq r_3 \neq k$, $k=1,\dots, N_{\rm ind}$; $r_1,r_2,r_3$ are integer random variables distributed as $\sim \mathcal{U}[1, N_{\rm ind}]$, and $F_{\rm mut} \in [0, 2]$ is the parameter representing the mutation scale factor. 

%-------------------------------------------------
\subsubsection{Crossover}
%-------------------------------------------------
The crossover vector is created from individual and mutation vectors following the rule:
\begin{equation}\label{eq:crossover}
\psi_{ik} = 
\begin{cases}
\nu_{ik}       & \quad \text{if } rand \in [0, 1] \leq F_{\rm cr} \text{ or } i = {r_k}\\
\iota_{ik}     & \quad \text{if } rand \in [0, 1] > F_{\rm cr} \text{ and } i \neq {r_k} 
\end{cases} 
\end{equation}
where $rand \sim \mathcal{U}\in [0, 1]$; $r_k$ is an integer $\sim \mathcal{U}[1, N_{\rm dim}]$ and $F_{\rm cr}\in [0, 1]$ is the crossover factor, one of the input parameters of the algorithm.

%-------------------------------------------------
\subsubsection{Selection}
%-------------------------------------------------
The population of individuals of the next generation is selected by the following rule:
\begin{equation}\label{eq:selection}
\boldsymbol{\iota}_{k}^{\textsc{g}} = 
\begin{cases}
\boldsymbol{\psi}_{k}       & \quad \text{if } f(\boldsymbol{\psi}_{k}) < f(\boldsymbol{\iota}_{k})\\
\boldsymbol{\iota}_{k}     & \quad \text{otherwise } 
\end{cases} 
\end{equation}

Notice that, in order to select the next generation, the fitness function must evaluate both the individuals and the crossover vectors, which reflects in the computational complexity of the algorithm.

After the execution of DE procedure $N_{\rm gen}$ times, the best individual $\boldsymbol{\iota}$ corresponds to the detected (estimated) symbol $\tilde{\bf x}_{\textsc{de}}[n]$ using the DE-aided detector in the MIMO-OFDM problem.  

\begin{algorithm}
\caption{ DE -- Differential Evolution.}\label{algo:DE}
\begin{algorithmic}[1]
\small
\State{ Input parameters: \, $F_{\rm cr}, F_{\rm mut}, N_{\mathrm{ind}}, N_{\mathrm{gen}},[\boldsymbol{\iota}_{1} \dots \boldsymbol{\iota}_{N_{\rm ind}}]$}
\For {$1 \to  N_{\mathrm{gen}}$}
\State{Mutation, eq. \eqref{eq:mutation}, $k = 1, \dots, N_{\rm ind}$}
\State{Crossover, eq. \eqref{eq:crossover}, $i=1,\dots,N_{\rm ind}; k = 1, \dots, N_{\rm ind}$ }
\State{ Select new individuals, eq. \eqref{eq:selection}, $k = 1, \dots, N_{\rm ind}$}
\EndFor
\State{Output: \, best individual $\boldsymbol{\iota}$}
\end{algorithmic}
\end{algorithm}

%======================================
\subsection{Hybrid Detectors} \label{subsec:initial}
%======================================
 To improve performance with a marginal increment on the computational complexity of the sub-optimal MIMO-OFDM  detectors, two efficient hybrid linear-heuristic algorithms are proposed and evaluated in the sequel. Starting from an initial solution provided by MMSE linear detector, a heuristic approach is applied in the subsequent stage aiming to improve the BER performance. In such hybrid configuration, the initial population/swarm in DE/PSO is generated adding random numbers with Gaussian distribution $\mathcal{N}(0,1)$ to the initial solution \cite{Storn1997}.  

In this work, different initial guess-solution are considered and numerical simulation are discussed under the perspective of the {\it performance-complexity tradeoff}. For that, numerical simulation results relating performance improvements and complex reduction are pointed out. Three different initializations have been considered herein:
\begin{enumerate}
\item {\it Random initialization}: initial positions (in the PSO) and population (DE) are generated using random variables uniformly distributed inside the search space.
\item {\it Hybrid approach}: two different initial points  are performed, which are provided by linear detectors MF and MMSE, while the respective symbol is considered as one variable input to the heuristic algorithms. 
\item {\it Perturbation on the MF/MMSE solutions}: the initial position of particles and initial population of individuals are obtained adding random Gaussian variables $\mathcal{N}(0,1)$ \cite{Storn1997} to the initial solution provided by MF/MMSE detector. 
\end{enumerate}

The influence of those points on the BER performance and complexity of the algorithm are explored in section \ref{sec:simulation_results}.

%=============================================
\section{Numerical Results} \label{sec:simulation_results}% (fold)
%============================================
Throughout this section, MIMO-OFDM systems are simulated considering realistic scenarios and different symbol detection. Specifically, linear, evolutionary heuristic and linear-heuristic detectors performance subject to spatial antenna correlation effect has been compared using BER and rates of convergence for heuristic and hybrid detector approaches. Moreover, for the heuristic-based  MIMO-OFDM detectors, the calibration of input parameters is conducted for each heuristic algorithm and respective hybrid approaches and the convergence reduction is appointed. After finding the best input parameter for each heuristic-based detector, the performance of the PSO and DE detectors are compared with hybrid approaches, namely PSO-MF, PSO-MMSE, DE-MF and DE-MMSE considering correlation between antennas; the performance of hybrid approaches are evaluated considering different number of iterations. Finally, the computational complexity of the algorithms are compared in terms of number of operations.

Table \ref{tab:mimo_OFDM} summarizes the simulation setup adopted in this work. Moreover, for a fair comparison, equal power allocation (EPA) was deployed throughout the transmitting antennas.

\begin{table}[!htb]
\caption{MIMO-OFDM simulation parameters. \label{tab:mimo_OFDM}}
\centering
\begin{tabular}{l l}    \hline
\textbf{Parameter}		&\textbf{Value} \\\hline\hline
\multicolumn{2}{c}{OFDM} \\
\hline
System Bandwidth, BW	&	20MHz     \\
Constellation 	&   4-QAM   	\\
Delay spread, $\tau_\textsc{rms}$ &     64ns   \\
\# Subcarriers, $N$      &       64\\
\hline
 \multicolumn{2}{c}{MIMO} \\
 \hline
\# Antennas, $N_t\times N_r$       & $4\times4$\\
Spatial correlation index & $\rho \in [0;\,\, 0.5; \,\, 0.9]$\\
MIMO-OFDM detectors & MF, ZF, MMSE, PSO, DE, PSO-MF, \\
                & PSO-MMSE, DE-MF, DE-MMSE\\
                Power allocation strategy & EPA \\
                \hline
                \multicolumn{2}{c}{Channel} \\
                \hline
                Type & NLOS Rayleigh channel\\
                CSI knowledge & perfect\\
                \hline
\multicolumn{2}{c}{ Heuristic Detectors Setup} \\
\hline
                Population size $N_{\rm pop} = N_{\rm ind}$ & {40} \\
                Search Space    &    [-1; 1] \\
\bottomrule
 \end{tabular}
  \end{table}

%-------------------------------------------------------
\subsection{Input Parameter Calibration for Heuristic-aided MIMO-OFDM Detectors}%-------------------------------------------------------
\begin{figure*}[!htbp]
 \centering
 %-----------------------------------------------------------------------
 \subfloat[Calibration: varying parameters and evaluating performance.
 ]{%
\includegraphics[width=.49\textwidth]{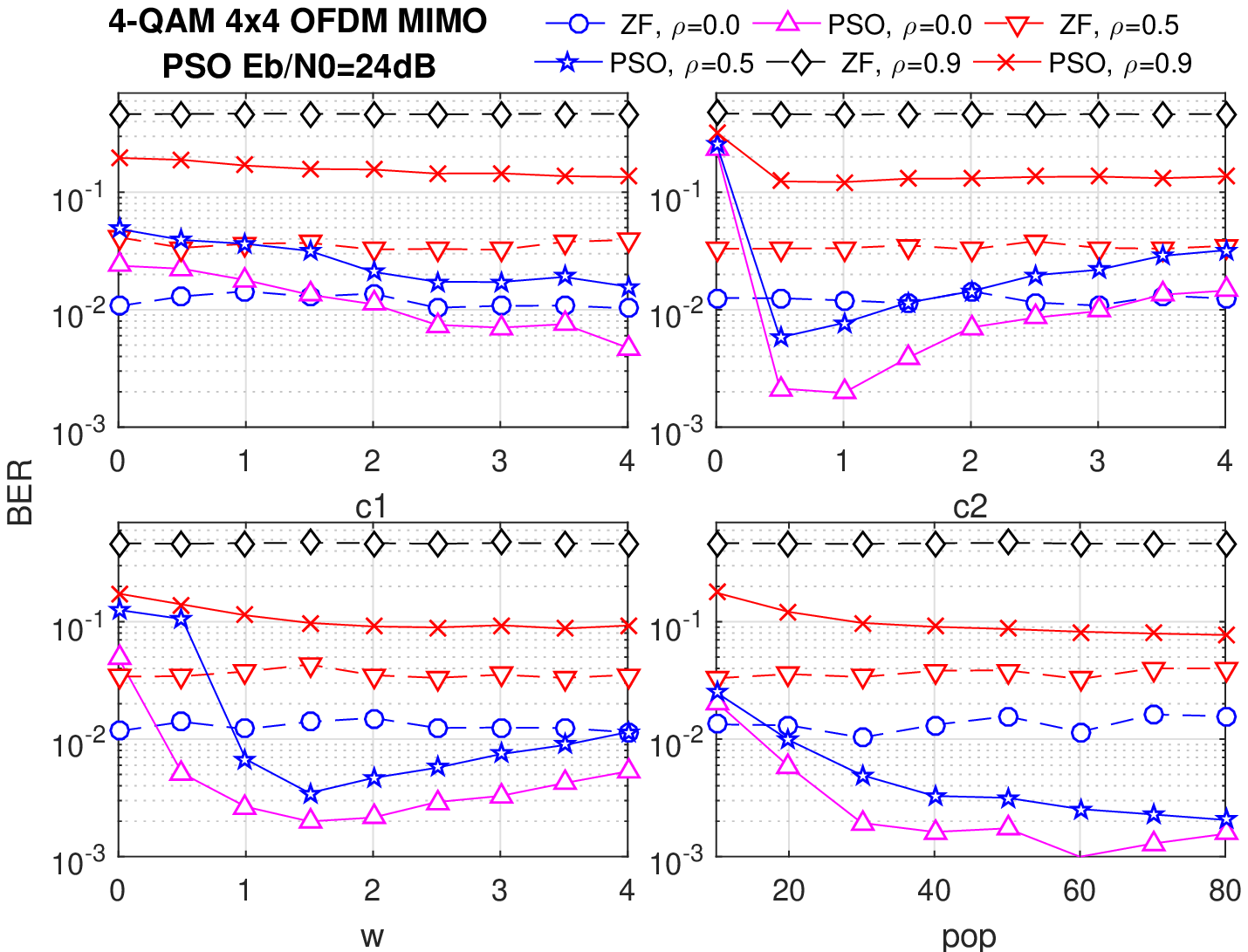}
}
\hfill
%-----------------------------------------------------------------------
 \subfloat[Calibration of input parameter of PSO-MF algorithm.
 ]{%
	\includegraphics[width=.49\textwidth]{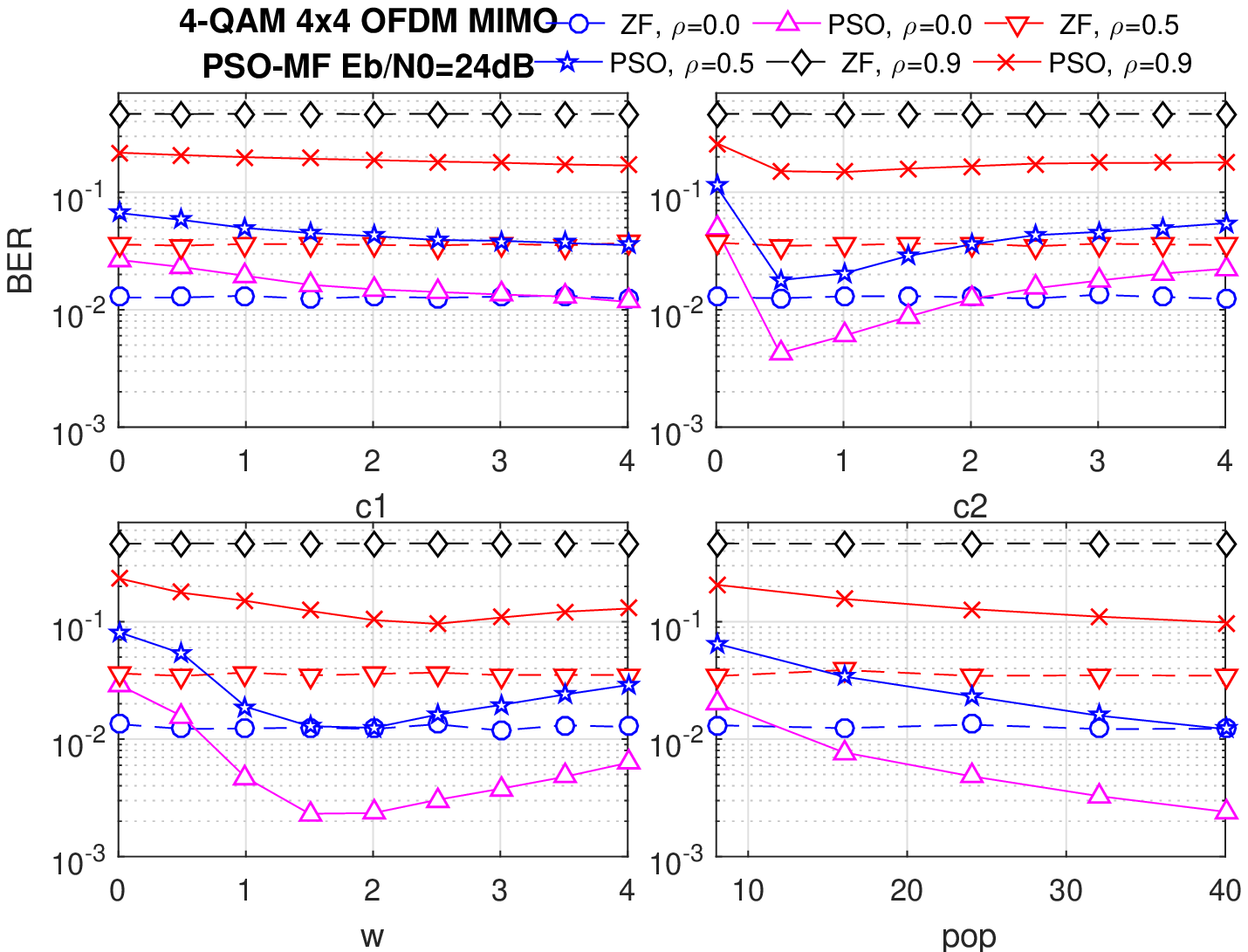}
}
\\
%-----------------------------------------------------------------------
\subfloat[Calibration of input parameters considering PSO-MMSE algorithm.
]{%
	\includegraphics[width=.48\textwidth]{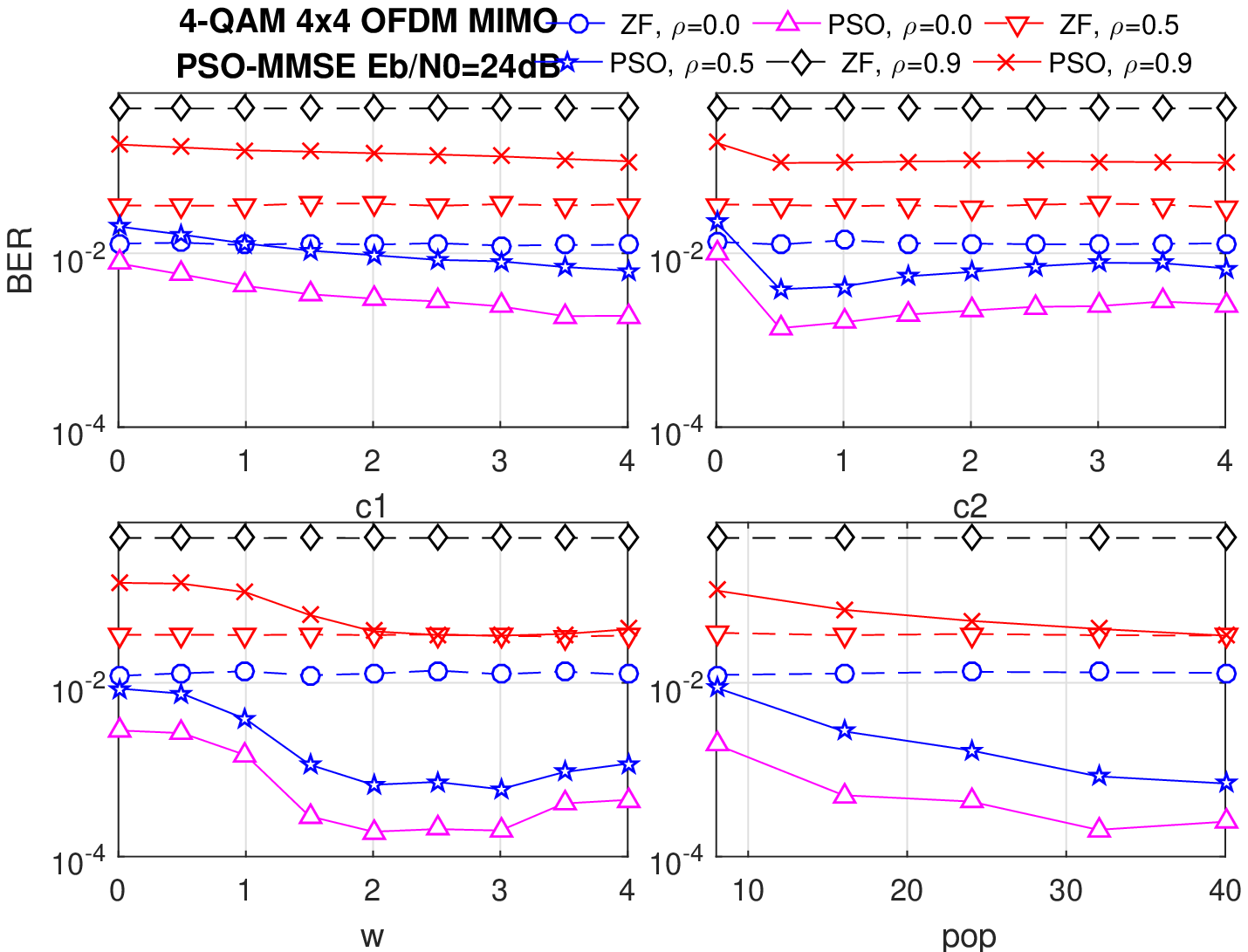}
}
\hfill
%-----------------------------------------------------------------------
\subfloat[{Convergence analysis for 4-QAM, $4 \times 4$ MIMO-OFDM with PSO detector considering different values of $E_b/N_0$. }
\label{fig:pso_convergence_4x4} 
]{%
\includegraphics[width=0.49\textwidth]{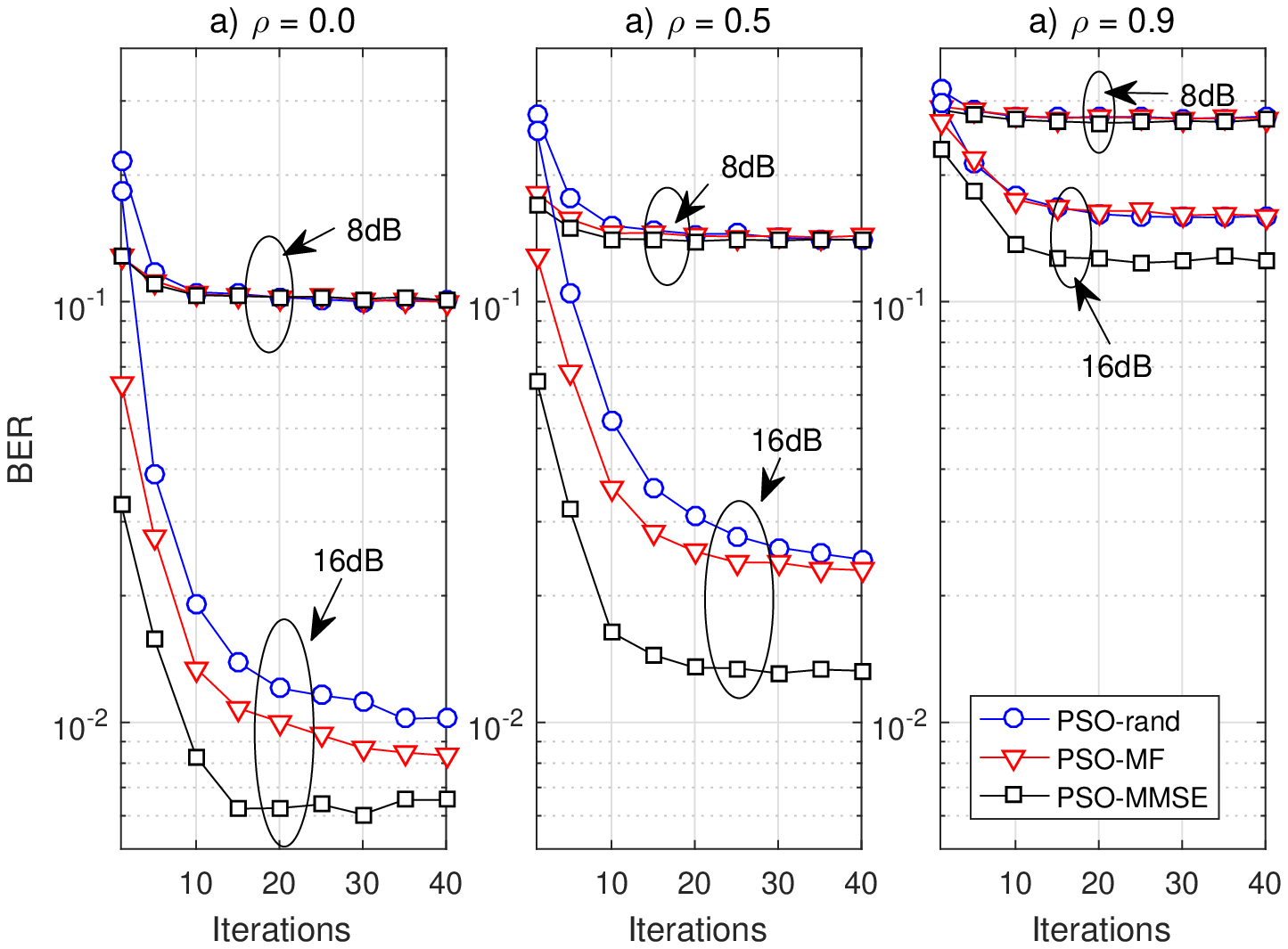}
}
\\
\caption{ Calibration of input parameters values for 4-QAM $4 \times 4$ MIMO-OFDM PSO detection problem operating under medium-high SNR and different spatial correlation indexes. } 
    \label{fig:pso_calib_4x4}
\end{figure*}

\begin{figure*}[!htbp]
 \centering
 %--------------------------------------------------------------------
 \subfloat[Calibration of input parameters of DE with uniformly random initialization.
 ]{%
\includegraphics[width=.49\textwidth]{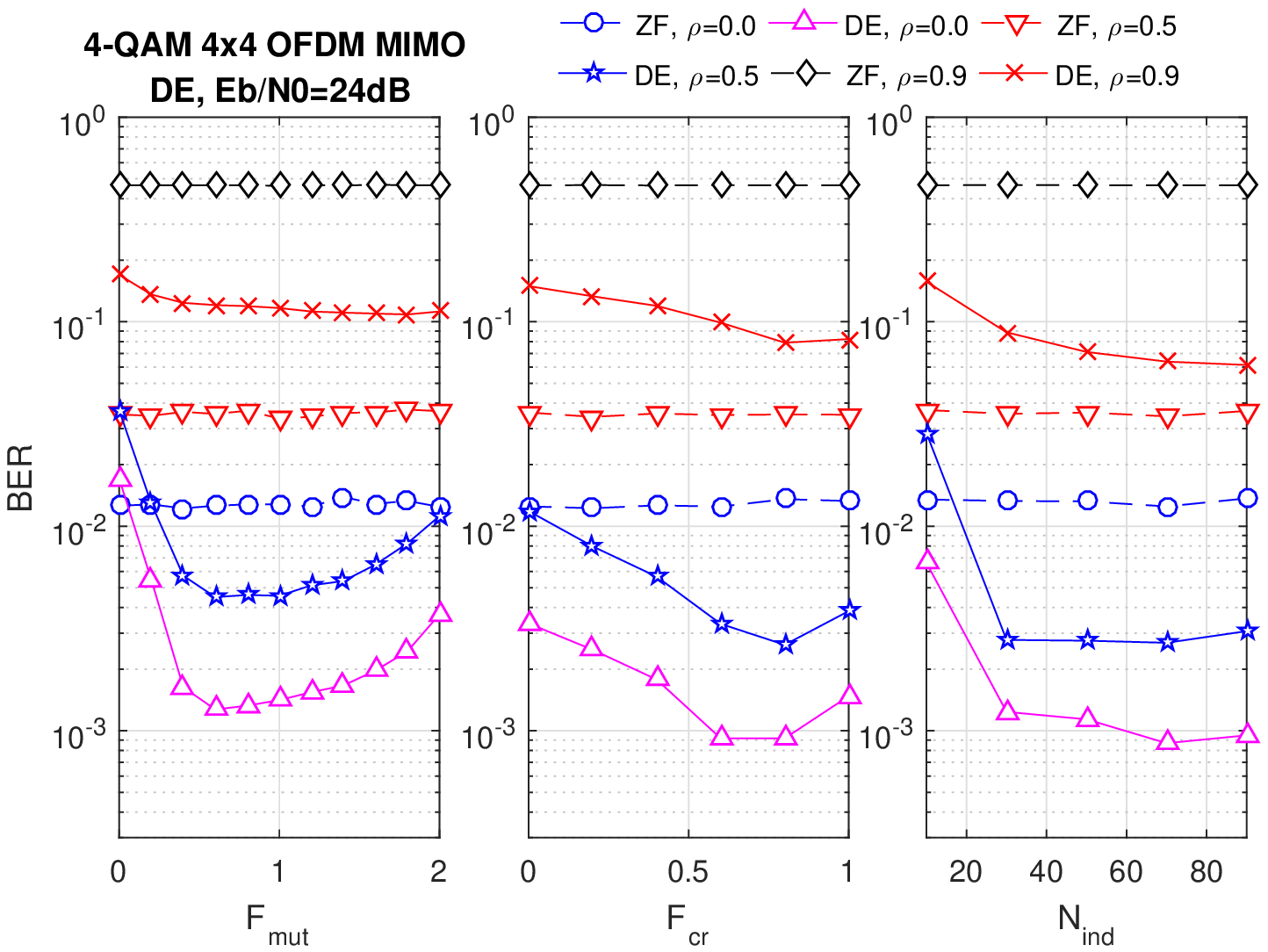}
}
\hfill
%--------------------------------------------------------------------
\subfloat[Calibration of input parameters of DE with MF initialization.
 ]{%
\includegraphics[width=.49\textwidth]{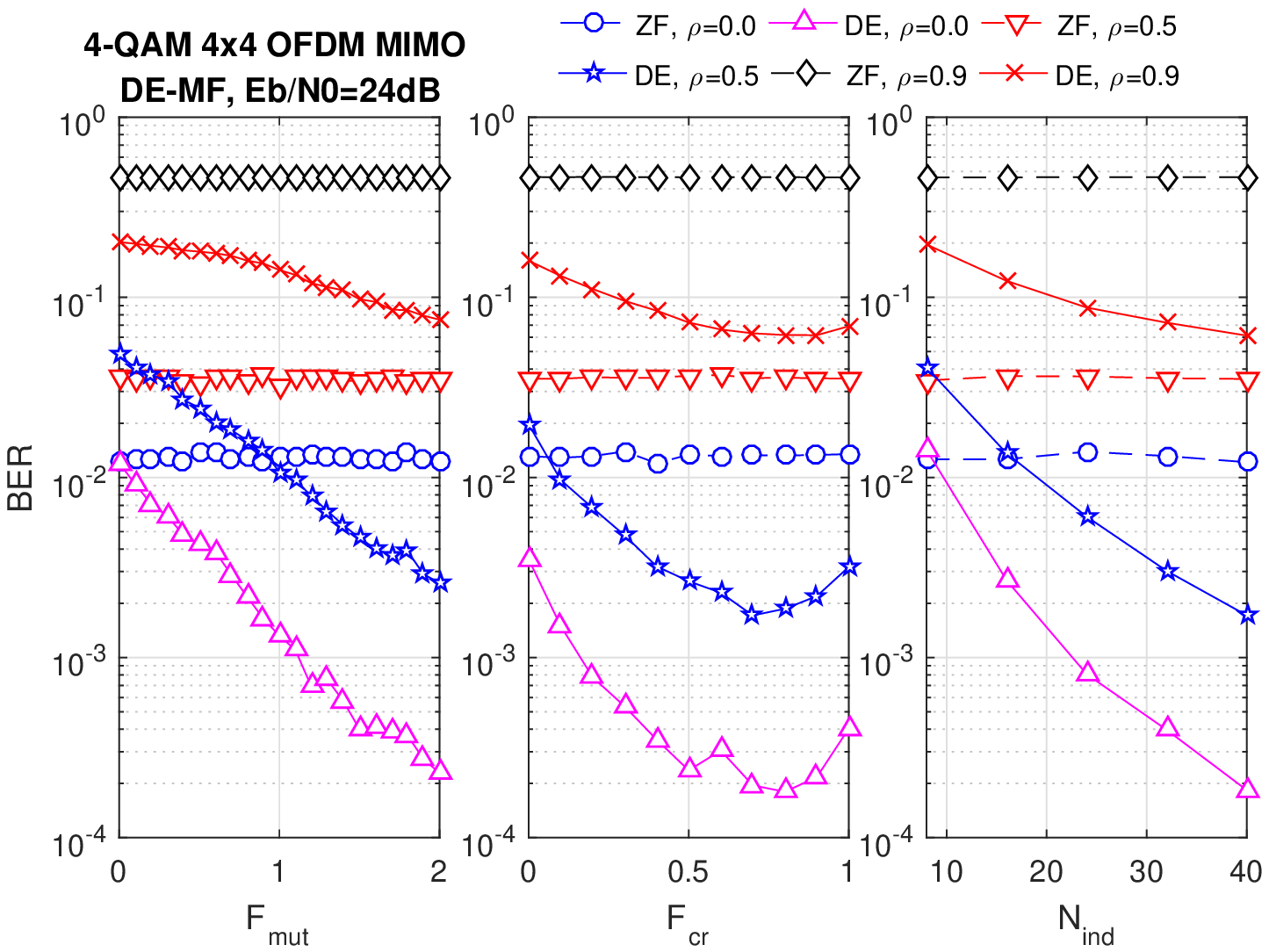}
}
\\
%--------------------------------------------------------------------
\subfloat[Calibration of input parameters of DE with MMSE initialization.
 ]{%
\includegraphics[width=.49\textwidth]{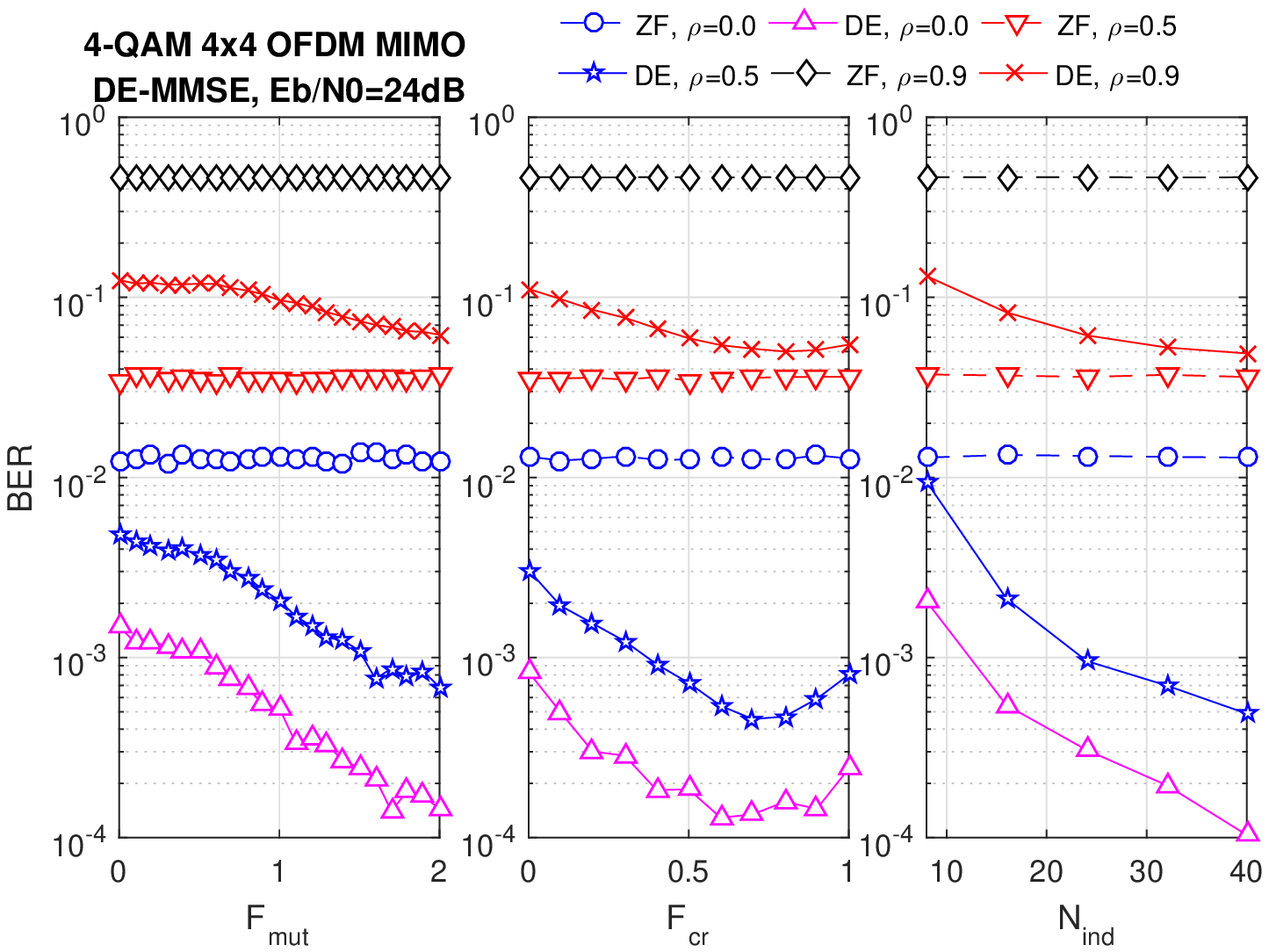}
}
\hfill
%--------------------------------------------------------------------
\subfloat[Convergence of DE-aided detector for MIMO-OFDM systems for different spatial correlation values. 
\label{fig:de_convergence4x4_4QAM}
 ]{%
\includegraphics[width=.49\textwidth]{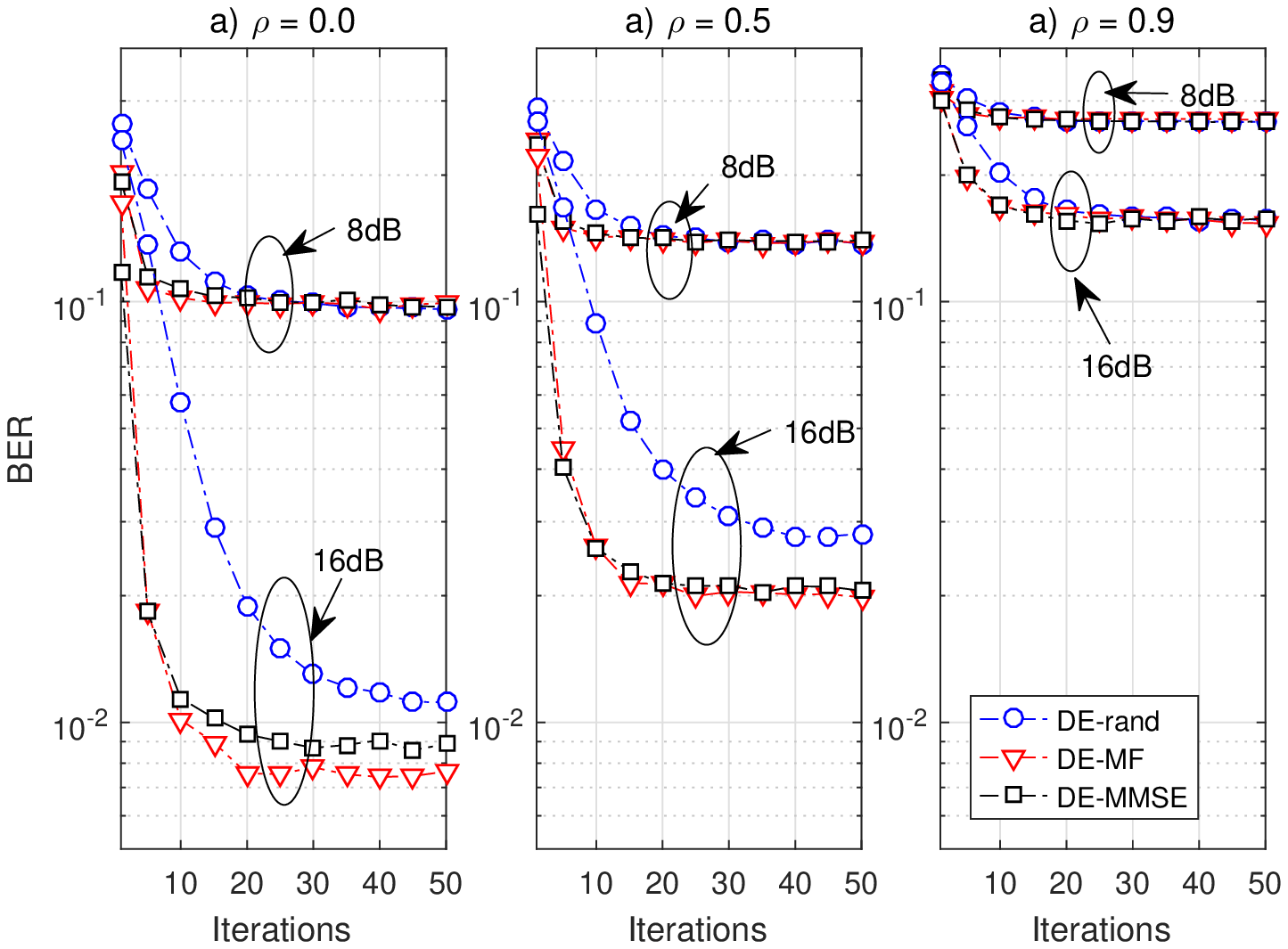}
}
\\
\caption{ Calibration of input parameters of DE heuristic applied to MIMO-OFDM detection for different values of correlation. } 
    \label{fig:deVarParRound1-2}
\end{figure*}

As different parameters may influence in the convergence properties of the heuristic algorithms, they were obtained numerically using the following procedure \cite{Marinello_2012}. Considering a set of start parameters, one by one is varied and the one that provides the lowest BER is considered in the variation of next parameter. The illustration of the procedure executed for PSO algorithm is presented in Fig. \ref{fig:pso_calib_4x4} and for DE algorithm in Fig. \ref{fig:deVarParRound1-2}, considering different values of spatial correlation and different initial points discussed in details in Subsection \ref{subsec:initial}. Observe that different initializations result in different initial parameters, which is more evident in the parameter $F_{\rm mul}$ for random and MF/MMSE initializations. Looking at the convergence in Fig. \ref{fig:pso_convergence_4x4}, one can notice that with MF and MMSE initialization, the number of iterations until convergence is reduced in comparison with random initialization case and consequently the complexity of the algorithm; as the $E_b/N_0$ value increases, more iterations are required. The start and final values after the calibration procedure for both PSO and DE heuristic-based detectors are summarized in Table \ref{tab:inputParametersPso} and \ref{tab:inputParametersDe}. 

\begin{table}[htbp]
  \centering
  \caption{Input parameters of PSO after calibration, considering $E_b/N_0=24{dB}$, different initial points and spatial correlation.}
  \label{tab:inputParametersPso}%
    \begin{tabular}{rl}
    \toprule
    {\bf Parameter} & {\bf Value}\\
    \midrule
	    {$N_{\rm iter}^{\rm start}$}    & [100; 20]\\
	    $c_1^{\rm start}$   & 2 \\
	    $c_2^{\rm start}$   & 2 \\
	    $w^{\rm start}$   & 1 \\    
    \hline
    %-------------------------------------------------------------
		%Iterations 
		{$N_{\rm iter}^{\rm rand}$}    & {100} \\
	    %Cognitive factor 
	    $c_1^{\rm rand}$   & 4 \\
	    %Social factor 
	    $c_2^{\rm rand}{(\rho)}$   & {1 ($0$) \,\,0.5 ($0.5$) \,\, 1 ($0.9$)} \\
	    %Inertia 
	    $w^{\rm rand}{(\rho)}$   &{1.5 ($0$) \,\,1.5 ($0.5$) \,\, 3.5 ($0.9$)} \\
	\hline
	%-------------------------------------------------------------
	    {$N_{\rm iter}^{\textsc{mf}}$}    & $\in [5; 25]$ \\
	    $c_1^{\textsc{mf}}$   & 4 \\
	    $c_2^{\textsc{mf}}{(\rho)}$   & {0.5 ($0$) \,\,0.5 ($0.5$) \,\,1 ($0.9$)} \\
	    $w^{\textsc{mf}}{(\rho)}$   &1.5 ($0$) \,\,{2 ($0.5$) \,\, 2.5 ($0.9$)} \\
    \hline
	%-------------------------------------------------------------
	    {$N_{\rm iter}^{\textsc{mmse}}$}    & $\in [5; 25]$ \\
	    $c_1^{\textsc{mmse}}(\rho)$   & 3.5($0$) \,\, 4($0.5$)\,\,  4($0.9)$\\
	    $c_2^{\textsc{mmse}}{(\rho)}$   & {0.5 ($0$) \,\,0.5 ($0.5$) \,\,0.5 ($0.9$)} \\
	    $w^{\textsc{mmse}}{(\rho)}$   &2 ($0$) \,\,{3 ($0.5$) \,\, 3 ($0.9$)} \\
    %------------------------------------------------------------
    \bottomrule
    \end{tabular}%
\end{table}%

\begin{table}[htbp]
  \centering
  \caption{ Input parameters of DE algorithm after calibration considering $E_b/N_0=24dB$, different initial points and spatial correlation.}
  \label{tab:inputParametersDe}%
    \begin{tabular}{rl}
    \toprule
    {\bf Parameter} & {\bf Value}\\
    \midrule
		$N_{\rm gen}^{\rm start}$  & [100; 20] \\
		$F_{\rm mut}^{\rm start}$ & 1 \\
		$F_{\rm cr}^{\rm start}$     & 0.5 \\
	\hline
	%-------------------------------------------------------------
		%\# generation   
		$N_{\rm gen}^{\rm rand}$  & {100} \\
		%Crossover factor 
		$F_{\rm cr}^{\rm rand}{(\rho)}$     & {0.6 ($0$) \,\,0.6 ($0.5$) \,\, 0.8 ($0.9$)} \\
		%Mutation factor
		$F_{\rm mut}^{\rm rand}(\rho)$ & {0.6 ($0$) \,\,0.8 ($0.5$) \,\, 1.8 ($0.9$)} \\
	\hline
	%-------------------------------------------------------------
	$N_{\rm gen}^{\textsc{mf}}$  & $\in [5; 25]$ \\
        $F_{\rm mut}^{\textsc{mf}}(\rho)$ & {2 ($0$) \,\,2 ($0.5$) \,\, 2 ($0.9$)} \\
        $F_{\rm cr}^{\textsc{mf}}{(\rho)}$     & {0.8 ($0$) \,\,0.7 ($0.5$) \,\, 0.9 ($0.9$)} \\
    \hline
	%-------------------------------------------------------------
$N_{\rm gen}^{\textsc{mmse}}$  & $\in [5; 25]$ \\
$F_{\rm mut}^{\textsc{mmse}}(\rho)$ & {1.7 ($0$) \,\,2 ($0.5$) \,\, 2 ($0.9$)} \\
	    $F_{\rm cr}^{\textsc{mmse}}{(\rho)}$     & {0.6 ($0$) \,\,0.7 ($0.5$) \,\, 0.8 ($0.9$)} \\
    %-------------------------------------------------------------
    \bottomrule
    \end{tabular}%
\end{table}%

%-------------------------------------------------------
\subsection{Performance Analysis}
\begin{figure}[ht]
 \centering
 \subfloat[Performance of hybrid algorithm PSO-MMSE.\label{fig:hybridPso}
 ]{%
\includegraphics[width=.49\textwidth]{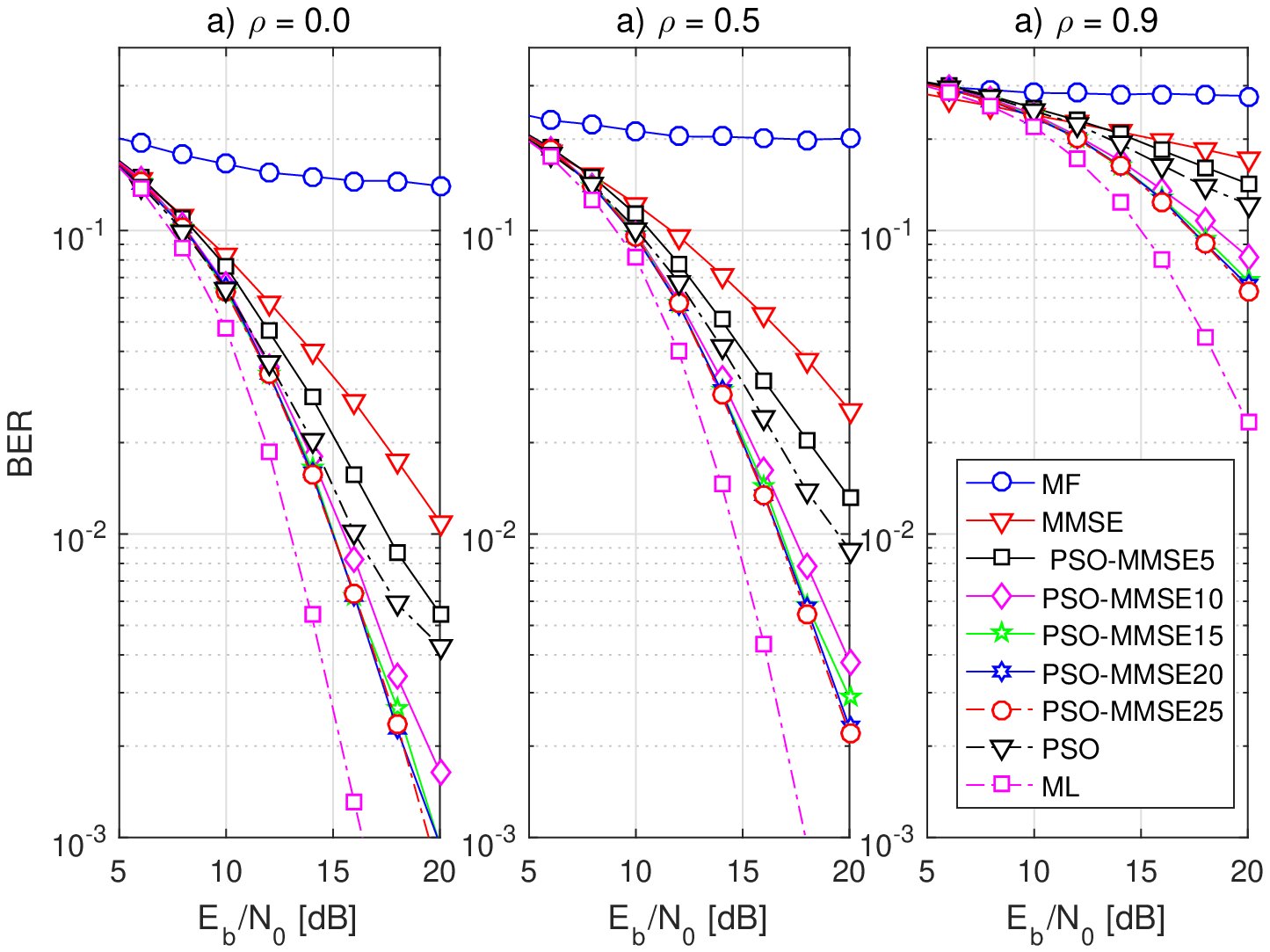}}
\hfill
%------------------------------------------------------
\subfloat[{Performance of hybrid algorithm DE-MMSE.  
\label{fig:hybridDe}} ]{%
\includegraphics[width=.48\textwidth]{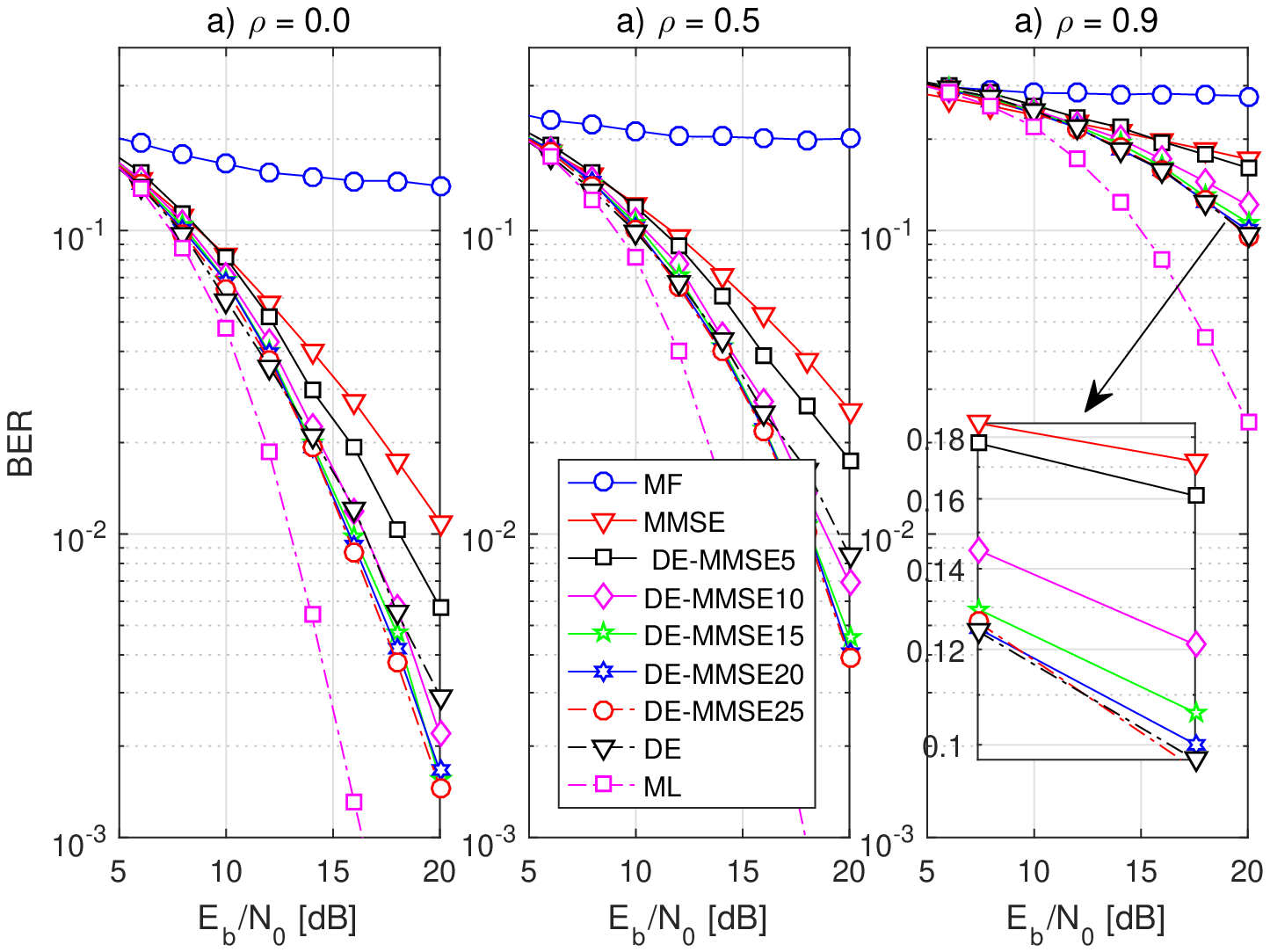}}
\\
\caption{Performance of the MMSE-hybrid algorithm considering ULA with different values of $E_b/N_0$, spatial correlation and increasing number of iterations. } 
    \label{fig:performance_hybrid_4x4_4QAM}
\end{figure}

\begin{figure}[ht]
 \centering
 \subfloat[Performance of hybrid algorithm PSO-MF.
 \label{fig:hybridPsoMf}]{%
\includegraphics[width=.49\textwidth]{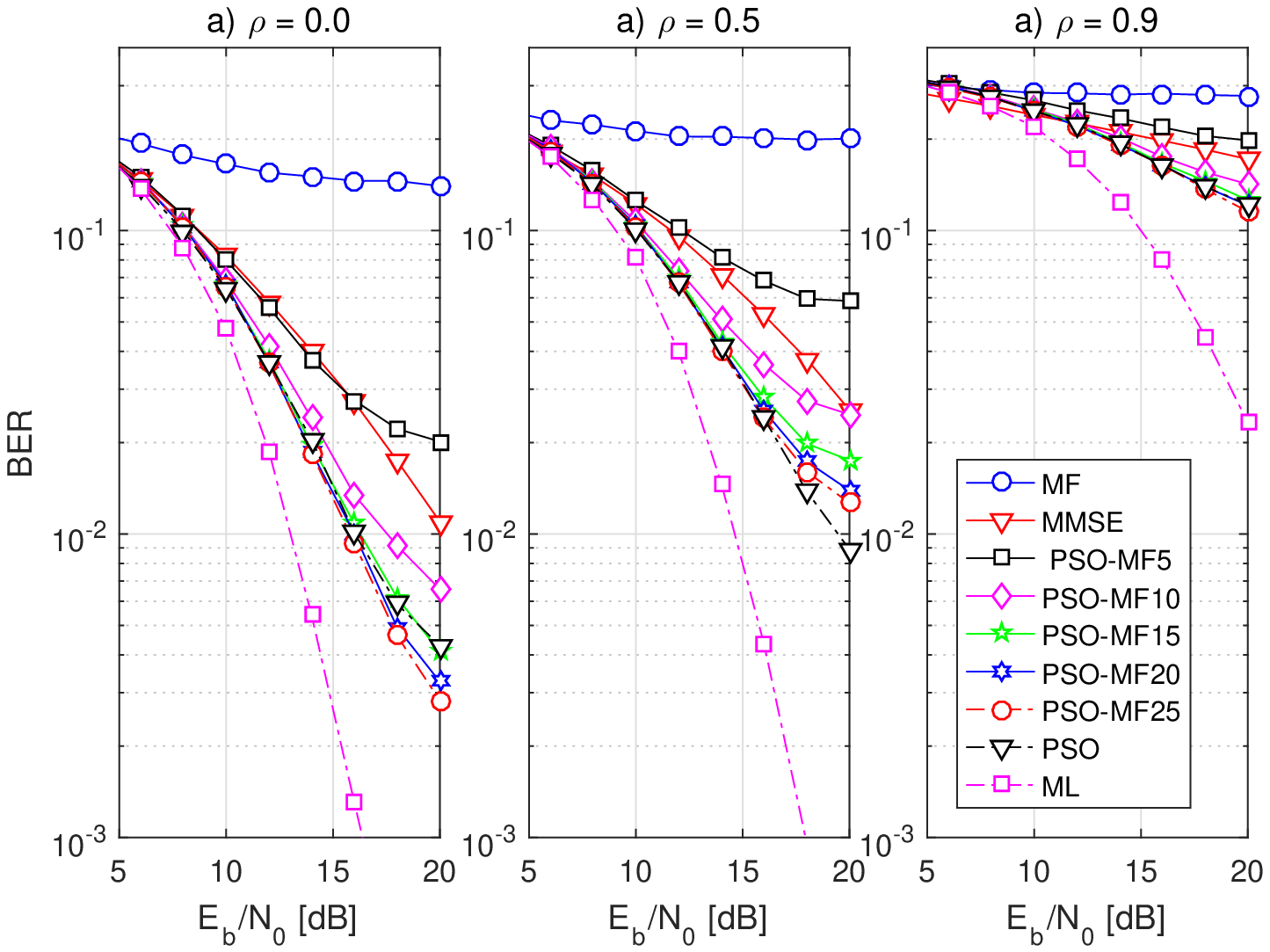}}
\hfill
%------------------------------------------------------
\subfloat[{Performance of hybrid algorithm DE-MF.  
\label{fig:hybridDeMf}} ]{%
\includegraphics[width=.48\textwidth]{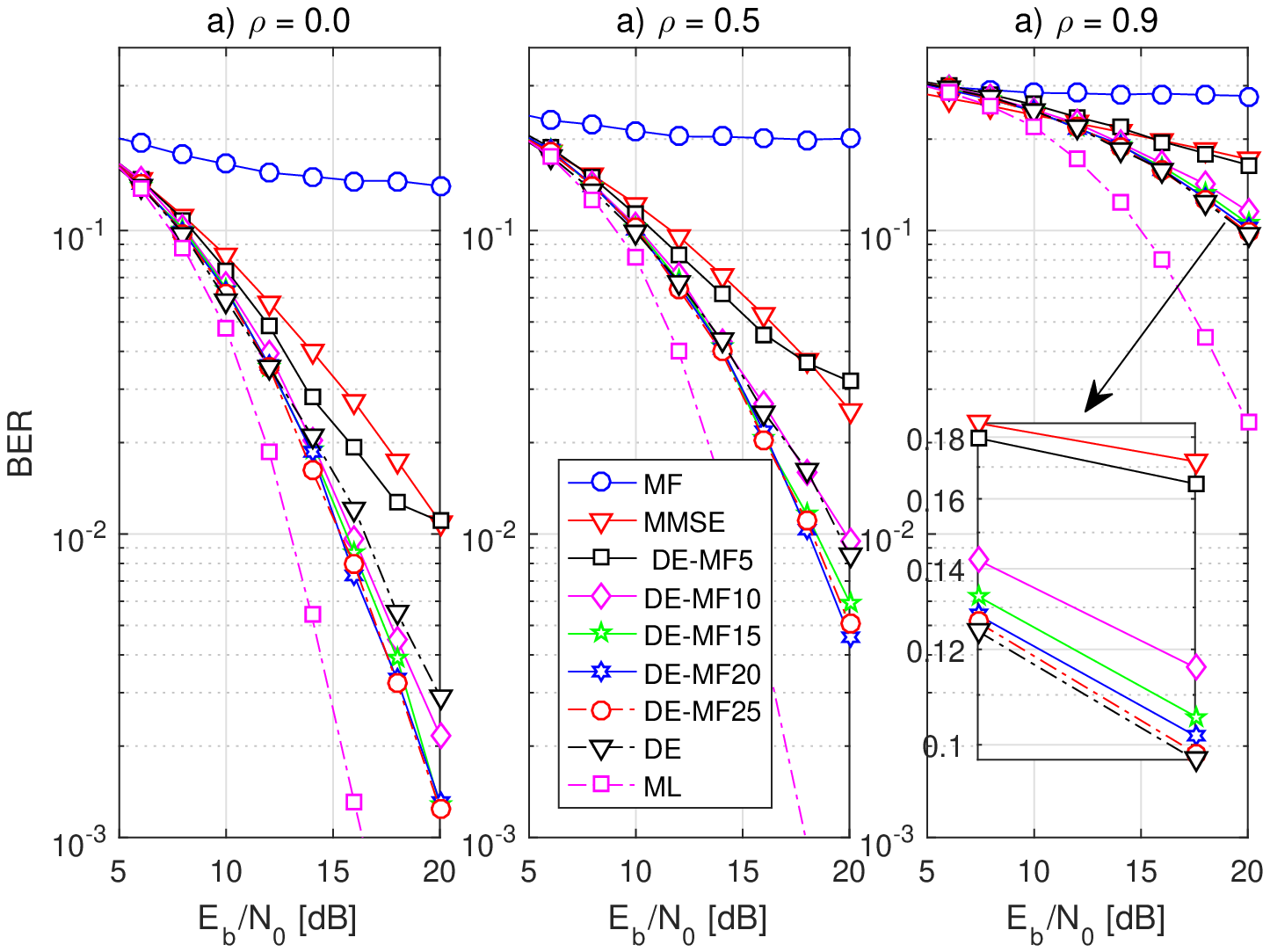}}
\\
\caption{Performance of the MF-hybrid algorithm considering ULA with different values of $E_b/N_0$, spatial correlation and increasing number of iterations.} 
    \label{fig:performance_MF_4x4_4QAM}
\end{figure}

%-------------------------------------------------------
After input parameters calibration, the BER performance of the heuristic and hybrid MIMO-OFDM detectors were numerically obtained. In Fig.\ref{fig:hybridPso} and \ref{fig:hybridDe}, the initial solution provided by the MMSE detector is considered. We observe that, as the number of iterations increase, the MMSE solution is refined and after 15 iterations, the improvement in BER performance becomes marginal for both algorithms DE-MMSE and PSO-MMSE.  
In \ref{fig:hybridPsoMf} and \ref{fig:hybridDeMf}, a similar behavior is observed. We note that the initial point influences the performance of PSO-based detectors: indeed, the PSO-MMSE provides better results in terms of BER than PSO-MF, but this effect is marginal for DE-MF and DE-MMSE, where similar performance is achieved after 15 iterations.

\begin{figure}[!htb]
\centering
\hspace{-3mm}\includegraphics[width=.5\textwidth]{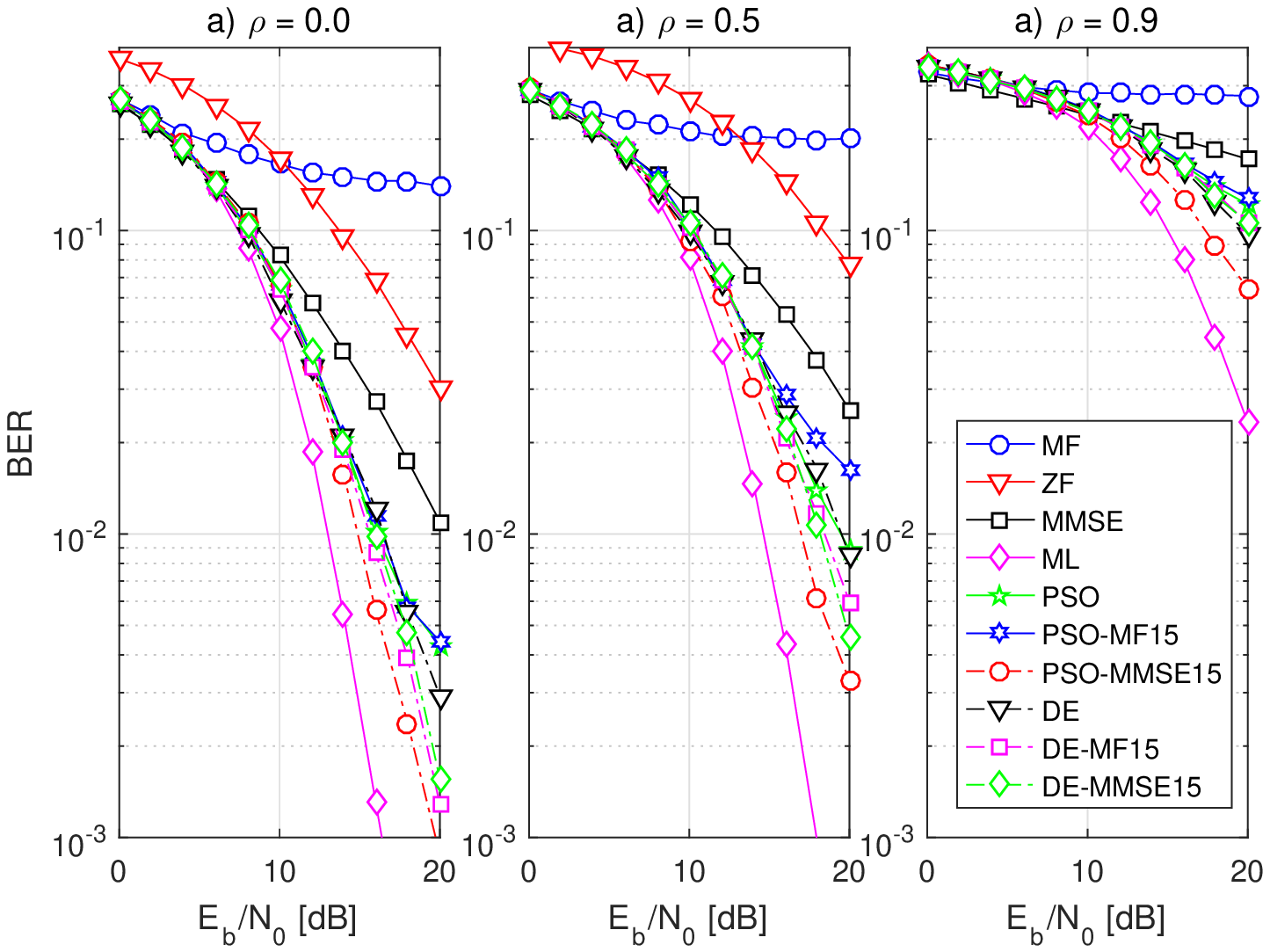}
\vspace{-5mm}
\caption{BER performance for 4-QAM, $4 \times 4$ linear array (ULA) antennas MIMO-OFDM for different detectors under different values of spatial correlation and SNR. } 
\label{fig:mimo_ofdm_correlation}
\end{figure}

In Fig.\ref{fig:mimo_ofdm_correlation}, the performances of linear, heuristic and hybrid MIMO-OFDM detection approaches are compared. We observe that PSO-MMSE provides the nearest ML performance, and that the hybrid approaches provide similar or better approaches than conventional heuristics. For highly correlated scenarios, the overall performance is worsened. For PSO-MMSE, the gain in performance is evident in contrast to other linear and heuristic detectors.

In general, spatial correlation degrades considerably the performance of all the studied detectors. However, hybrid heuristic-linear MIMO-OFDM detectors are suitable choices for MIMO systems operating under low or even moderate antenna correlation.

%-------------------------------------------------------
\subsection{Complexity Analysis}\label{ref:complexity}
%-------------------------------------------------------
 To analyze the complexity of the detection algorithms, the number of \textsc{flop}s among real numbers are considered. The \textsc{flop}s are described as floating point addition, subtraction, multiplication or division operations \cite{golub2012}. In this evaluation, Hermitian operator and \texttt{if} conditional step  were disregarded. In practice, some platforms use hardware random number generators, where an electric circuit provides random numbers generation, and so the \textsc{flop}s cost to generate random numbers was also ignored.

\begin{figure}[!htbp]
\centering
\includegraphics[width=0.49\textwidth]{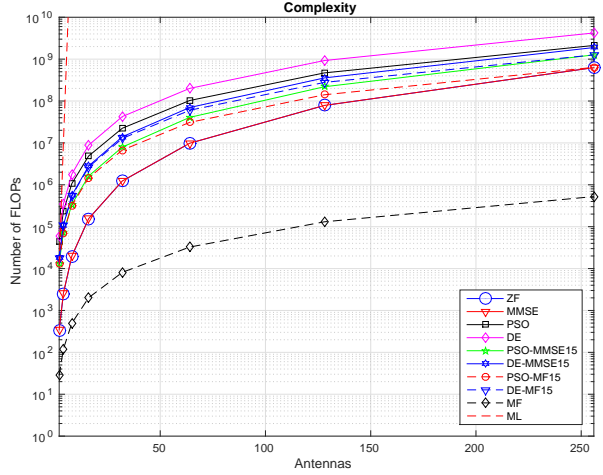}
\vspace{-5mm}
\caption{MIMO-OFDM Complexity considering an increasing number of antennas for linear, heuristic and hybrid detectors in a point-to-point scenario; $N_t = N_r, N_{\rm dim} = 2N_t, N_{\rm pop} = N_{\rm ind} = 5\cdot N_{\rm dim}, \mathcal{I} = 50, \mathcal{I}_{\rm hyb} = 15$. } 
\label{fig:flops}
\end{figure}

Table \ref{tab:referenceFlops} describes the number of \textsc{flop}s needed for the main operations considered  herein, while in Table \ref{tab:flopLinearHeuristicDetectors}, the full complexity expressions ($\Upsilon$) for the analyzed MIMO-OFDM detectors are shown. In Fig.\ref{fig:flops}, the complexity is described considering typical values, {\it i.e.}, $N_{\dim} = 2N_t; N_t = N_r; N_{\rm ind} = N_{\rm pop} = 5\cdot N_{\dim}$ and admitting the number of iterations up to the convergence obtained previously through simulations, as shown in Fig. \ref{fig:pso_convergence_4x4}, \ref{fig:de_convergence4x4_4QAM} for the heuristic algorithms and for the hybrid algorithm in Fig. \ref{fig:performance_hybrid_4x4_4QAM}  and \ref{fig:performance_MF_4x4_4QAM}. 

From Table \ref{tab:flopLinearHeuristicDetectors}, it can be observed that DE algorithm requires more \textsc{flop}s than PSO since it evaluates $2N_{\rm pop}$ times the fitness function per iteration in eq. \eqref{eq:selection} for individuals and crossover vectors. The complexity between the linear detectors are almost the same, differing from each other by an scalar-matrix multiplication and matrix-matrix sum in eq. \eqref{eq: ZFD} and eq. \eqref{eq:MMSED}. Moreover, observing the hybrid heuristic-linear MIMO-OFDM detector in Fig. \ref{fig:performance_hybrid_4x4_4QAM} and \ref{fig:performance_MF_4x4_4QAM}, the improvement in performance starts to stagnate around 15 iterations, and so $\mathcal{I}_{\rm hyb}=15$ has been considered as the number of iterations of the hybrid algorithm to attain the best performance-complexity tradeoff. 

\begin{table}[htbp]
  \centering
  \caption{Number of \textsc{flop}s, considering vector and matrices ${\bf w}\in \mathbb{R}^{q \times 1}, {\bf A}\in\mathbb{R}^{m\times q}, {\bf B}\in\mathbb{R}^{q\times p}, {\bf C}\in\mathbb{R}^{m\times p}, {\bf D}\in\mathbb{R}^{q\times q}$.  }
  \label{tab:referenceFlops}%
    \begin{tabular}{p{0.31\textwidth}p{0.08\textwidth}}
    \toprule
    \bf Operation & \# \textsc{flop}s\\
    \midrule
    Square root $\sqrt{.}$  & {8} \\
    Norm-2, $\sqrt{{\bf w}^T{\bf w}}$	& $2n-1 + 8$\\
    Matrix-vector multiply {\bf Aw} & {$m(2q-1)$} \\
    {Matrix-matrix  multiply ${\bf AB}$} & {$mp(2q-1)$} \\
    Matrix multiply-add ${\bf AB+C}$ & $2mpq$ \\
    Matrix inversion with LU factorization of {\bf D} \cite{BoydNumerical} & $2/3q^3 + 2q^2$ \\
    \bottomrule
    \end{tabular}%
\end{table}%

Heuristic detection algorithms produce better BER performance at the cost of an incremental computational complexity compared with linear detectors ZF and MMSE, mainly due to the population/swarm size (around $5$ to $10\cdot N_{\rm dim}$) and number of iterations necessary to attain convergence. In order to reduce the complexity, both hybrid linear-heuristic algorithms combing MF/MMSE and evolutionary-heuristic techniques were analyzed. The PSO-MF provides computational complexity near the linear approaches for $N_t=256$ antennas. PSO-MMSE has similar computational complexity than DE-MF.

Although linear MMSE and heuristic algorithms have slightly more computational complexity than other linear approaches, there is also improvement in BER performance. Moreover, evolutionary heuristics may be more flexible to be implemented in hardware. Parallelization, the possibility to deal with non-differentiable and nonlinear functions \cite{Storn1997} and the possibility to truncate the number of iterations to achieve different performance-complexity trade-offs in scenarios that do not require very low levels of BER, for example with MF hybrid, may be good choices for real applications.

\begin{table}[!htbp]
\centering
\caption{Number of \textsc{flop}s per subcarrier for the MIMO-OFDM detectors, with $\underline{\bf H} \in \mathbb{R}^{2N_r \times 2N_t}$, $\underline{\bf y} \in \mathbb{R}^{2N_r \times 1}$, $N_{\rm dim} = 2N_t$.}
\label{tab:flopLinearHeuristicDetectors}%
\tiny
 \begin{tabular}{p{0.18\textwidth}p{0.27\textwidth}} 
\toprule
  \bf   Detector & \bf Number of Operations \\
    \midrule
    $\Upsilon_\textsc{MF}(N_t, N_r)$    & $2N_t(4N_r - 1)$ \\
    %--------------------------------------------
    {$\Upsilon_\textsc{ZF}(N_t, N_r)$}     & {$\dfrac{16}{3}N_t^3 + 4N_t^2 + 32N_t^2 N_r + 4N_tN_r - 2N_t$ } \\
    %--------------------------------------------
    {$\Upsilon_\textsc{MMSE}(N_t, N_r)$}     & {$\dfrac{16}{3}N_t^3 + 8N_t^2 + 32N_t^2N_r + 4N_tN_r$ } \\
    %--------------------------------------------
    {$\Upsilon_\textsc{PSO}(N_t, N_r, N_{\rm pop}, \mathcal{I})$}     & {$N_{\rm pop} \mathcal{I} ( 8N_tN_r + 20N_t + 4N_r + 7)$ } \\
    %--------------------------------------------
    {$\Upsilon_\textsc{DE}(N_t, N_r, N_{\rm ind}, \mathcal{I})$}     &  { $N_{\rm ind} \mathcal{I} (16N_tN_r + 12N_t + 8N_r + 14) $}\\
    %--------------------------------------------
    {$\Upsilon_\textsc{PSO-MMSE}(N_t, N_r, N_{\rm pop}, \mathcal{I}_{\rm hyb})$}     & {$\Upsilon_\textsc{PSO}(N_t, N_r, N_{\rm pop}, \mathcal{I}_{\rm hyb}) + \Upsilon_\textsc{MMSE}(N_t, N_r)$ } \\
    %--------------------------------------------
    {$\Upsilon_\textsc{DE-MMSE}(N_t, N_r, N_{\rm ind}, \mathcal{I}_{\rm hyb})$}     &  {$\Upsilon_\textsc{DE}(N_t, N_r, N_{\rm ind}, \mathcal{I}_{\rm hyb}) + \Upsilon_\textsc{MMSE}(N_t, N_r)$ }\\
    %--------------------------------------------
    $\Upsilon_\textsc{PSO-MF}(N_t, N_r, N_{\rm pop}, \mathcal{I}_{\rm hyb})$     & $\Upsilon_\textsc{PSO}(N_t, N_r, N_{\rm pop}, \mathcal{I}_{\rm hyb}) + \Upsilon_\textsc{MF}(N_t, N_r)$ \\
    %--------------------------------------------
    $\Upsilon_\textsc{DE-MF}(N_t, N_r, N_{\rm ind}, \mathcal{I}_{\rm hyb})$     &  $\Upsilon_\textsc{DE}(N_t, N_r, N_{\rm ind}, \mathcal{I}_{\rm hyb}) + \Upsilon_\textsc{MF}(N_t, N_r)$\\
    %--------------------------------------------
    \hline
    $\Upsilon_\textsc{ML}(N_t, N_r, \mathcal{M})$     &   $\mathcal{M}^{2N_t} ( 8N_tN_r + 4N_r + 7) $\\
    %--------------------------------------------
 \bottomrule
  \multicolumn{2}{l}{$\mathcal{I}:$ \# iterations for conventional algorithms }\\
   \multicolumn{2}{l}{$\mathcal{I}_{\rm hyb}:$ \# iterations for the hybrid algorithm}\\
\end{tabular}
\end{table}

%=======================================================
\section{Conclusions}\label{sec:conclusions}
%======================================================
Extensive simulations were deployed and suitable evolutionary heuristic PSO and DE input parameters calibration were chosen numerically aiming to find suitably and of practical interest solutions for the MIMO-OFDM detection problem. Hybrid approaches considering MF and MMSE as initial solutions have been also considered, where the linear initial solution is improved while the number of iterations of heuristic algorithms reduced.

Among the analyzed  MIMO-OFDM detectors, the hybrid PSO-MMSE provided the near-ML performance for the considered scenarios, {\it i.e.} $\rho = 0$ (uncorrelated), $\rho = 5$ and $\rho = 0.9$. However, the BER performance has demonstrated be sensible to the initialization. For PSO-MF, the performance was similar to conventional PSO, with the advantage of reduced number of iterations until convergence. For DE, almost the same BER performance was achieved using MF and MMSE.

In terms of complexity, ZF and MMSE require almost the same number of \textsc{flop}s, although MMSE requires some statistical knowledge of the channel condition. Among the heuristic detectors, DE requires more \textsc{flop}s in comparison with the PSO, mainly because the number of fitness function evaluations is higher, since in DE it is calculated for the $\boldsymbol{\iota}_k$ and $\boldsymbol{\psi}_k, k = 1,\dots, N_{\rm ind}$ per iteration of the algorithm, in comparison to $N_{\rm pop}$ per iteration with PSO (in the simulations, $N_{\rm pop}=N_{\rm ind}$).

To improve the complexity-performance tradeoff, this work proposed and evaluated two linear-heuristic hybrid algorithms suitable to solve the MIMO-OFDM detection problem. Starting from a solution obtained from the MMSE and MF linear detectors, the DE and PSO heuristics were executed in order to further improve the BER performance while they were able to improve substantially the performance-complexity tradeoff even under low and medium spatial correlation scenarios. Numerical simulations have demonstrated that with both hybrid algorithms, the number of iterations required to the convergence is reduced, achieving similar and slightly better performance in the DE and PSO-hybrid detectors when compared to the conventional DE and PSO.

\section*{Acknowledgment}
This work was supported in part by the National Council for Scientific and Technological Development (CNPq) of Brazil under Grants 130464/2015-5 (Scholarship) and 304066/2015-0 (Researcher grant), in part by Araucaria Foundation, PR, under Grant 302/2012 (Research) and by Londrina State University - Paraná State Government, Brazil.

\end{document}